\documentclass{article}
\usepackage{spconf,amsmath,graphicx}
\usepackage[skip=1pt]{caption}
\usepackage{bm}
\usepackage{makecell}
\usepackage{cite}

\title{Poisson Noise Removal Using Multi-Frame 3D Block Matching}
\name{}
\address{}
\name{Kireeti Bodduna$^{1,2}$ and Joachim Weickert$^1$}
\address{
\begin{minipage}[b]{0.5\linewidth}
{\centering
$^1$Mathematical Image Analysis Group, \\
Saarland University, \\ 66041 Saarbr{\"u}cken, Germany.\\
\{bodduna,weickert\}@mia.uni-saarland.de \\}
\end{minipage}
\begin{minipage}[b]{0.51\linewidth}
\centering
$^2$Saarbr{\"u}cken Graduate School \\ of Computer Science, \\
Saarland University, \\ 66041 Saarbr{\"u}cken, Germany.
\end{minipage}}

\begin{document}
%
\maketitle
\begin{abstract}
The 3D block matching (BM3D) filter belongs to the state-of-the-art 
techniques for eliminating additive white Gaussian noise from 
single-frame images. There exist four multi-frame extensions of BM3D 
as of today. In this work, we combine these extensions with a variance 
stabilising transformation (VST) for eliminating Poisson noise. Our 
evaluation reveals that the extension which retains the original 
noise model of the noisy images and additionally has a  
comprehensive connectivity of 2D and temporal image information 
at both pixel and patch levels, gives the best results. 
Additionally, we find a surprising change in performance 
of one the four extensions due to the specific application 
of the VST. Finally, we also introduce a simple low-pass filtering 
as a preprocessing step for the best performing extension. 
This can give rise to a significant additional improvement of 0.94 dB 
in the output according to the peak signal to noise ratio.
\end{abstract}
\begin{keywords}
Multi-frame denoising, non-local patch methods, 
Poisson noise, 3D block matching 
\end{keywords}
\section{Introduction}
\label{sec:intro}
Image acquisition through CCD/CMOS sensors is dominated by 
Poisson noise distribution \cite{rooms2003PoissonDenoising}. 
Astronomical imaging \cite{borkowski2010Astronomy}, 
medical imaging \cite{rodrigues2008DenoisingPoisson}, 
and electron microscopy \cite{jinschek2008EMPoisson} are some 
of the specific applications where we encounter Poisson noise. 
There are two categories of imaging techniques that are designed 
to eliminate Poisson noise: In one of the categories, additive white 
Gaussian noise (AWGN) elimination algorithms 
\cite{DFKE07, boulanger2008Denoising, zhang2008WaveletsPoisson, 
fryzlewicz2012HaarPoisson, portilla2003Denoising} are combined with 
variance stabilising transformations (VST) \cite{anscombe1948PoissonTrans, 
makitalo2011optimal, makitalo2011ClosedForm}. 
The other category comprises of algorithms that are designed for 
eliminating Poisson noise directly \cite{lefkimmiatis2009BayesianPoisson, 
luisier2010WaveletPoisson, willett2003PlateletsPoisson}. 
Among all these techniques, 3D block matching (BM3D) \cite{DFKE07} 
when combined with a VST, gives the best denoising results in most 
cases \cite{makitalo2011optimal}.   

All the above mentioned techniques concentrate on denoising 
single-frame images. There has not been much investigation 
in the field of multi-frame image denoising. In applications 
such as electron microscopy, CT imaging and multi-spectral imaging, 
we encounter situations where multiple images of the same 
scene can be acquired. Further processing is required for 
fusing the information from all the images to one single image. 
Several approaches \cite{blu2007sure, hasan2018denoising, 
boulanger2010patch, dong2015low, 
Buades2009, tico2008multi, buades2010MultiFrame, fang2012sparsity,
luisier2009MultiFrame, zhang2009MultiFrame} that have been designed 
to remove both AWGN and Poisson noise from multi-frame image datasets, 
are inspired from algorithms that are originally designed for 
single-frame image denoising. 

Non-local patch-based methods \cite{DFKE07, LBM2013, Lebrun2012} 
are among the best performing methods for single-frame 
image denoising. Efforts in the direction of optimally extending 
non-local patch based methods to multi-frame elimination 
of signal dependent noise \cite{Buades2009, 
buades2010MultiFrame}, made use a of hybrid filtering scheme.
In particular, depending on the temporal standard deviation after 
registration, they use a combination of values computed through 
simple averaging of the frames and primitive spatio-temporal non-local 
patch-based methods. We have revived the research in this direction 
in \cite{bodduna2019MultiFrame}, by studying all possible 
spatio-temporal extensions of 3D block matching in the multi-frame 
scenario. In particular, we have evaluated our proposed extensions and 
existing ones for eliminating AWGN. 
We have learnt from this evaluation that using multi-frame extension 
ideas which retain the noise model and simultaneously 
enhance the connectivity between frames, give superior results.
The optimal performance of such ideas in the AWGN layout, 
motivates us to also analyse their capability in the multi-frame Poisson 
noise scenario.   

{\bf Our Contribution.} There exist four extensions of BM3D 
for multi-frame AWGN removal \cite{bodduna2019MultiFrame}. 
In the present work, we perform a comparative 
evaluation of these extensions for eliminating Poisson noise 
from multi-frame image datasets. For this purpose, we combine all 
four extensions with the Anscombe transform and the closed-form 
approximation of the exact unbiased inverse Anscombe transform. 
The difference in the qualitative performance according to 
peak-signal-to-noise (PSNR) ratio, between the 
closed-form approximation and the exact unbiased inverse is 
small, but the former is faster \cite{makitalo2011ClosedForm}. 
Hence, we use the closed-form approximation for computing the inverse 
variance stabilising transformation. Additionally, we introduce 
a preprocessing step for the best performing extension after the 
evaluation, in the form of simple low-pass filtering. This 
extra step significantly improves the denoising output.

{\bf Paper Structure.} In Section \ref{sec:modelling}, we 
introduce our novel framework that uses various extensions of 
BM3D for eliminating Poisson noise from multi-frame image datasets. 
In Section \ref{sec:expAndDisc}, we present our experimental results. We 
also provide explanations behind the respective positions of different 
extensions in the ranking. Finally, in Section \ref{sec:concAndOutlook}, 
we present a summary of our work and give an outlook to future 
work. 


\section{Modelling of denoising algorithm}
\label{sec:modelling}

There are two trivial and two non-trivial extensions among the 
four existing extensions of BM3D for multi-frame AWGN removal
\cite{bodduna2019MultiFrame}. 
In Sections \ref{trivial} and \ref{non-trivial} we 
review these four extensions. In Sections \ref{variance_stab} 
and \ref{low-pass} we explain in detail the exact procedure that 
makes optimal use of these four extensions for multi-frame 
Poisson noise elimination.  

\subsection{Trivial Extensions of BM3D}
\label{trivial}
{\bf BM3D-1:} In the first trivial extension \cite{bodduna2019MultiFrame}, 
we first average all the frames after registering 
them. In the second step, we use the original BM3D method \cite{DFKE07} 
to denoise the averaged image. \\ \\
{\bf BM3D-2:} In the second trivial extension 
\cite{bodduna2019MultiFrame}, we first denoise every single frame 
using the original BM3D method. In the second step, we average 
all the denoised frames after registering them. 


\begin{table*}[t]
\setlength{\tabcolsep}{3pt}
\begin{minipage}{0.53\textwidth}
\begin{flushright}
\begin{tabular}{ l r  r r  r r r}
 \hline 
 Image & BM-1 & BM-2 & BM-3 & BM-M & 
 $\sigma$/ BM-M$_\sigma$\\
 \hline 

 House (1) & 18.29 & 22.73 & 22.94 & 24.28 & 95/ \textbf{24.74}\\
 House (2) & 21.19 & 26.02 & 25.53 & 27.11 & 100/ \textbf{27.28}\\
 House (3) & 23.15 & 27.29 & 26.49 & 28.13 & 110/ \textbf{28.32}\\  
 House (4) & 24.69 & 28.14 & 27.23 & 29.06 & 105/ \textbf{29.32}\\  
               \vspace{0.5em}
 House (5) & 25.97 & 28.87 & 27.91 & 29.74 & 120/ \textbf{29.92}\\
 Lena (1) & 19.00 & 24.54 & 23.90 & 25.33 & 195/ \textbf{25.88}\\
 Lena (2) & 21.64 & 26.53 & 25.82 & 27.31 & 200/ \textbf{27.50}\\
 Lena (3) & 23.43 & 27.67 & 26.77 & 28.44 & 220/ \textbf{28.62}\\ 
 Lena (4) & 24.84 & 28.33 & 27.41 & 29.13 & 215/ \textbf{29.24}\\  
  \vspace{0.5em}
 Lena (5) & 25.86 & 28.94 & 27.95 & 29.67 & 210/ \textbf{29.82}\\      
 Bridge (1) & 18.02 & 20.79 & 20.40 & 21.16 & 130/ \textbf{21.85}\\
 Bridge (2) & 19.85 & 21.93 & 21.50 & 22.36 & 145/ \textbf{22.88}\\
 Bridge (3) & 20.95 & 22.59 & 22.08 & 23.08 & 140/ \textbf{23.55}\\
 Bridge (4) & 21.73 & 23.04 & 22.48 & 23.55 & 145/ \textbf{24.00}\\  
   \vspace{0.5em}
 Bridge (5) & 22.26 & 23.38 & 22.80 & 23.89 & 145/ \textbf{24.33}\\   
  Peppers (1) & 20.48 & 24.75 & 23.97 & 25.53 & 170/ \textbf{26.10}\\ 
 Peppers (2) & 22.78 & 26.68 & 25.83 & 27.46 & 195/ \textbf{27.63}\\ 
 Peppers (3) & 24.49 & 27.69 & 26.77 & 28.41 & 205/ \textbf{28.54}\\ 
 Peppers (4) & 25.64 & 28.42 & 27.48 & 29.14 & 205/ \textbf{29.24}\\ 
 Peppers (5) & 26.50 & 28.89 & 27.94 & 29.57 & 205/ \textbf{29.67}\\
 \hline 
 \end{tabular}
\end{flushright}
\end{minipage}
\hspace{1em}
\vspace{1em}
\begin{minipage}{0.35\textwidth}
 \begin{tabular}{ r  r r  r r}
 \hline 
 BM-1 & BM-2 & BM-3 & BM-M & 
 $\sigma$/ BM-M$_\sigma$\\
 \hline 
   17.55 & 22.98 & 23.34 & 25.20 & 80/ \textbf{26.11}\\  
   20.51 & 26.41 & 25.79 & 27.93 & 90/ \textbf{28.26}\\ 
   22.67 & 27.77 & 26.73 & 29.26 & 105/ \textbf{29.58}\\  
   24.25 & 28.60 & 27.55 & 30.09 & 100/ \textbf{30.48}\\ 
  \vspace{0.5em}
   25.60 & 29.28 & 28.23 & 30.73 & 100/ \textbf{31.06}\\   
   18.16 & 24.78 & 24.12 & 25.95 & 165/ \textbf{26.91}\\ 
   20.98 & 26.92 & 26.07 & 28.24 & 180/ \textbf{28.71}\\ 
   22.92 & 28.05 & 26.99 & 29.41 & 195/ \textbf{29.75}\\ 
   24.42 & 28.75 & 27.64 & 30.13 & 195/ \textbf{30.42}\\  
  \vspace{0.5em}
   25.55 & 29.28 & 28.08 & 30.64 & 195/ \textbf{30.94}\\    
   17.41 & 20.90 & 20.53 & 21.46 & 115/ \textbf{22.53}\\   
   19.49 & 22.10 & 21.64 & 22.86 & 130/ \textbf{23.66}\\  
   20.77 & 22.72 & 22.18 & 23.61 & 135/ \textbf{24.32}\\   
   21.62 & 23.20 & 22.61 & 24.19 & 145/ \textbf{24.81}\\ 
  \vspace{0.5em}
   22.23 & 23.55 & 22.90 & 24.57 & 145/ \textbf{25.15}\\   
   19.65 & 24.99 & 24.20 & 26.22 & 160/ \textbf{27.07}\\  
   22.16 & 26.99 & 26.09 & 28.30 & 175/ \textbf{28.71}\\  
   23.99 & 28.02 & 26.95 & 29.31 & 180/ \textbf{29.60}\\ 
   25.23 & 28.78 & 27.65 & 30.03 & 175/ \textbf{30.25}\\ 
   26.14 & 29.24 & 28.12 & 30.47 & 185/ \textbf{30.69}\\ 
 \hline
\end{tabular}
\end{minipage}
 \vspace{1em}
 \captionof{table}{PSNR values after denoising 5-image (left)
 and 10-image (right) datasets with noise peaks varying from 1 to 5.}
 \label{table1}
\end{table*}

\subsection{Non-Trivial Extensions of BM3D}
\label{non-trivial}
{\bf BM3D-3:} The original BM3D method is a two-step process. Initially, 
for every reference patch considered, a corresponding 
3D group of patches is formed by searching for the most similar 
patches in the image with respect to the $L_2$ distance. 
In the first non-trivial extension BM3D-3 \cite{buades2005denoising,
tico2008multi, Buades2009, buades2010MultiFrame},
we consider a reference frame initially. For every reference 
patch in this frame we increase the search area for finding 
the most similar patches from this particular frame 
to all the frames. The further procedure is similar to the original 
BM3D method. \\ \\
{\bf BM3D-4:} The second non-trivial 
extension \cite{bodduna2019MultiFrame} is also a two-step method with 
three sub-steps in each step like the original BM3D method: \\
{\textit{Step 1.1 - Grouping.}} Unlike BM3D-3, here we consider reference 
patches from all the frames but not just one frame. 
A corresponding 3D group for every reference patch is then created by 
finding the most similar patches in all the frames using $L_2$ 
distance. We remove the threshold parameters for $L_2$ distance.
This is done because there is a risk of losing similar patches for 
high amplitudes of noise. However, we retain the parameters that 
specify the maximum number of patches in a 3D group. \\ 
{\textit{Step 1.2 - Filtering.}} We apply the following techniques  
on every obtained 3D group, in the same order: 2D bi-orthogonal 
spline wavelet transform, 1D Walsh-Hadamard transform, hard 
thresholding, 1D Walsh-Hardamard back transform, and a 2D 
bi-orthogonal spline wavelet back transform. \\ 
{\textit{Step 1.3 - Aggregation.}} Now, every pixel in every frame 
is denoised at least once after the second sub-step. The denoised 
versions of every pixel is present in 3D groups belonging to 
the frame the pixel belongs to, as well as in 3D groups from other 
frames. Thus, we first compute a weighted aggregation of all the denoised 
versions of every pixel within the frame in which it is present. 
This gives us as many initial denoised frames as there are input frames. 
To compute a final denoised frame after the first main step, we 
use the following equation for weighted aggregation across 3D 
groups in all frames:  
\begin{equation}
\label{equation}
\bm{u}^{\textrm{\textbf{basic}}}({\bm{x}}) = 
  \frac{\sum\limits_\ell \sum\limits_{P_\ell} 
  w_{P_\ell}^{\textrm{\textbf{hard}}} 
  \sum\limits_{Q \in P(\textrm{P}_\ell)} \chi_Q(\bm{x}) 
  \bm{u}^{\textrm{\textbf{hard}}}_{Q,P_\ell}(\bm{x})}
  {\sum\limits_l \sum\limits_{P_\ell} 
  w_{P_\ell}^{\textrm{\textbf{hard}}} 
  \sum\limits_{Q \in P(\textrm{P}_\ell)} \chi_Q(\bm{x}) }.
\end{equation}
Here, $\bm{x}$ is a position in the 2D image domain $Q$ and  
$\bm{u}^{\textrm{\textbf{basic}}}$ is the initial denoised image. 
The set of most similar patches to the reference patch 
$P_\ell$ belonging to frame $\ell$, are denoted using $\mathcal P (P_\ell)$. 
We have $\chi_Q(\bm{x}) = 1 $ if $\bm{x} \in Q$ and 0 otherwise,
for every patch $Q$ in the set $\mathcal P (P_\ell)$. 
The symbol $\bm{u}^{\textrm{\textbf{hard}}}_{Q,P_{\ell}}(\bm{x})$  
denotes the estimation of the value at pixel position $\bm{x}$,  
belonging to the patch $Q$,  derived after the hard thresholding  
(with coefficients $w_{P_\ell}^{\textrm{\textbf{hard}}}$)
of the reference patch $P_\ell$.\\ 
{\textit{Step 2.1 - Grouping.}}   
The same grouping strategy as in \textit{1.1} is employed by us,  
using the reference patches from the initial denoised frames 
computed after the first main step. \\ 
{\textit{Step 2.2 - Filtering.}} We execute the following 
techniques on every obtained 3D group and the 
corresponding initial noisy 3D group, in the same order: 2D discrete 
cosine transform, 1D Walsh-Hadamard transform, Wiener filtering on a 
combination of both the corresponding 3D groups, 1D Walsh-Hadamard 
back transform, and a 2D discrete cosine back transform. \\ 
{\textit{Step 2.3 - Aggregation.}} The same corresponding 
strategy as in \textit{1.3} is exploited by us to estimate the final 
denoised image. This aggregation strategy can be represented as:
\begin{equation}
\label{equation_2}
\bm{u}^{\textrm{\textbf{final}}}({\bm{x}}) = 
  \frac{\sum\limits_\ell \sum\limits_{P_\ell} 
  w_{P_\ell}^{\textrm{\textbf{wien}}} 
  \sum\limits_{Q \in P(\textrm{P}_\ell)} \chi_Q(\bm{x}) 
  \bm{u}^{\textrm{\textbf{wien}}}_{Q,P_\ell}(\bm{x})}
  {\sum\limits_l \sum\limits_{P_\ell} 
  w_{P_\ell}^{\textrm{\textbf{wien}}} 
  \sum\limits_{Q \in P(\textrm{P}_\ell)} \chi_Q(\bm{x}) }.
\end{equation}
Here $\textrm{\textbf{wien}}$ represents Wiener filtering
and $\bm{u}^{\textrm{\textbf{final}}}$ is the final denoised image. The 
rest of the symbols have the same meaning as in (\ref{equation}). 
Moreover, we can also represent the initial denoised image aggregated 
using BM3D-3 by \eqref{equation}. Choosing $\ell = 1$ 
in \eqref{equation} gives us the initial estimate of 
the single-frame BM3D algorithm. The final denoised image using 
BM3D-3 and single-frame BM3D can both be computed by selecting 
$\ell = 1$ in \eqref{equation_2}. 

We apply the particular filtering steps 1.2 and 2.2 in the Fourier 
domain because of better differentiation between signal and noise when 
compared to the Cartesian domain. The original work \cite{DFKE07} provides 
more specific details regarding the single-frame BM3D algorithm. 

In the first extension, we have a change in the noise 
distribution because of averaging the noisy frames.
In BM3D-2, we denoise every frame first, which is a 
sub-optimal solution since we have limited amount 
of signal in each of the frames. BM3D-3 avoids the above two 
modelling disadvantages by searching for 
similar patches in all the frames. However, we consider just 
one reference frame and this does not allow us to make use of the 
complete information. In the final extension BM3D-4, we consider 
every frame as the reference frame. This allows us to compute 
similar patches from all the frames in both the main steps, 
unlike BM3D-3. Such a modelling also allows us to perform a better 
aggregation for obtaining the final denoised image. 
This is due to the presence of more denoised versions of every pixel 
in the 2D image domain, when compared to the third extension.


\begin{figure*}[t]
  \centering
  \includegraphics[width=0.2\linewidth]
  {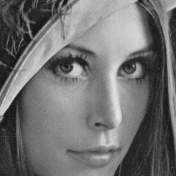}\hspace{0.1em}
  \includegraphics[width=0.2\linewidth]
  {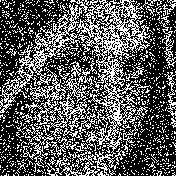}\hspace{0.1em}
  \includegraphics[width=0.2\linewidth]
  {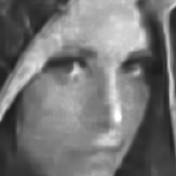}\hspace{0.1em}
  \vspace{0.2em}
  \includegraphics[width=0.2\linewidth]
  {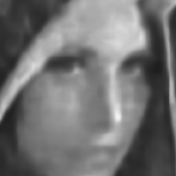}
  \includegraphics[width=0.2\linewidth]
  {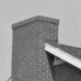}\hspace{0.1em}
  \includegraphics[width=0.2\linewidth]
  {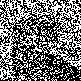}\hspace{0.1em}
  \includegraphics[width=0.2\linewidth]
  {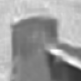}\hspace{0.1em}
    \vspace{0.2em}
  \includegraphics[width=0.2\linewidth]
  {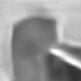}  	
  \includegraphics[width=0.2\linewidth]
  {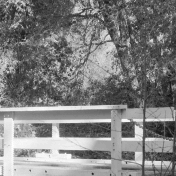}\hspace{0.1em}
  \includegraphics[width=0.2\linewidth]
  {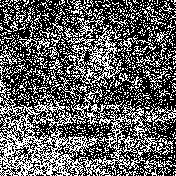}\hspace{0.1em}
    \vspace{0.2em}
  \includegraphics[width=0.2\linewidth]
  {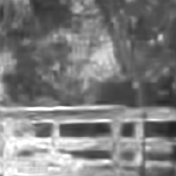}\hspace{0.1em}
  \includegraphics[width=0.2\linewidth]
  {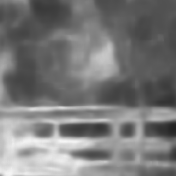}
  \includegraphics[width=0.2\linewidth]
  {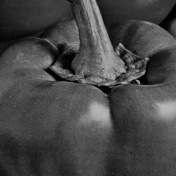}\hspace{0.1em}
  \includegraphics[width=0.2\linewidth]
  {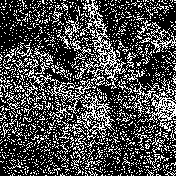}\hspace{0.1em}
  \includegraphics[width=0.2\linewidth]
  {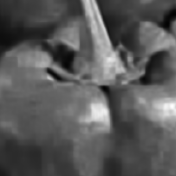}\hspace{0.1em}
  \vspace{0.2em}
  \includegraphics[width=0.2\linewidth]
  {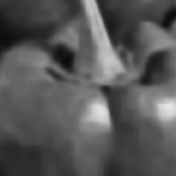} 
\caption{Results after denoising 5-image datasets with 
noise peak of 1.0. Top to bottom: Zoom into the Lena, House, 
Bridge, Peppers images. Left to right: Original, noisy, 
BM3D-4$_\sigma$, next best method.}
\label{fig:res}
\end{figure*}
\subsection{Denoising with Variance Stabilisation}
\label{variance_stab}
We carry out the following procedure for eliminating Poisson noise 
by using the various BM3D extensions mentioned above: 
All the noisy datasets undergo the Anscombe transformation for 
variance stabilisation, affine rescaling to [0,1] greyscale range, 
filtering using one of the four extensions, affine rescaling back 
to the original range, and finally the closed-form approximation of the 
exact unbiased inverse Anscombe transformation. 

In BM3D-1, we have applied the transformation on the averaged image while 
in all other extensions we have applied the VST directly on the initial
noisy images. One can find more details 
regarding the above general procedure and parameter selection in 
\cite{makitalo2011optimal, makitalo2011ClosedForm,hou2010comments}.

\subsection{Low-Pass Filtering}
\label{low-pass}
As already mentioned in Section \ref{sec:intro}, we use a simple low 
pass filter \cite{Frangakis2001} as a preprocessing step for the 
best performing extension. The following low-pass filter is applied on 
the noisy images directly (without variance stabilisation or affine 
rescaling), in order to make the process of finding similar patches that 
form a 3D group more robust:
\begin{equation*}
H(\bm{\hat{x}}) = \begin{cases}
1 &\text{$ \textrm{for} \ \ |\bm{\hat{x}}| < \frac{\sigma}{2}$},\\
 \textrm{exp} \left(\frac{-(\hat{x} - (\frac{\sigma}{2}))^2
 				-(\hat{y} - (\frac{\sigma}{2}))^2}
 				{2(\frac{\sigma}{2})^2} \right)
		&\textrm{for} \ \ |\bm{\hat{x}}| \ge \frac{\sigma}{2}.
\end{cases}
\end{equation*}
Here, $\bm{\hat{x}} := (\hat{x} , \hat{y})^\top$ 
represents the 2D frequency vector in 
the Fourier domain and $\sigma$ specifies the 
shape of the filter. The Gaussian-type decay of the 
filter coefficients is intended to reduce the transform-domain 
filtering artifacts. We use the low-pass filtered image
only for forming the 3D groups in the first main step and
nowhere else in the entire algorithm. 


\section{Experiments and Discussion}
\label{sec:expAndDisc}

In multi-frame denoising applications, the frames are first 
registered using motion registration algorithms before denoising 
them. In this work we want to specifically test the denoising 
capability of different extensions of BM3D. To this end, 
we assume that the frames are perfectly registered similar to 
\cite{bodduna2019MultiFrame}. This assumption 
makes even more sense because it has already been verified 
\cite{Buades2009, buades2010MultiFrame} that denoising 
methods (even without temporal support) are superior to 
averaging at regions where there is high temporal standard 
deviation after registration. Thus, verifying the performance of 
different denoising methods assuming perfect registration also 
covers the remaining case of less temporal standard deviation. 
Thereby, to create this experimental setting, 
we have added Poisson noise to the Lena, House, Peppers and 
Bridge\footnote{http://sipi.usc.edu/database/} images  with 
noise peaks varying from 1 to 5. Due to the signal dependent 
nature of Poisson noise, we have high noise amplitude when the 
intensity of the signal decreases. The noise peak value controls this 
trade-off. Lesser noise peak value indicates higher amount of noise.
For every image and every noise peak, we have created two datasets 
each with 5 and 10 realisations of noise.  

Dissolving the threshold parameters for $L_2$ distance while 
forming the 3D groups is advantageous to all the extensions. 
Hence, we have excluded this parameter in both the main steps of 
BM3D for all the four extensions. 

For BM3D-3, in the results which we showcase shortly, we present 
the best PSNR after every frame 
is considered as the reference frame. 

From Table \ref{table1} and Figure \ref{fig:res}, 
we can conclude that both visually and in 
terms of PSNR, the non-trivial extension BM3D-4 outperforms 
all other extensions significantly. 
The comprehensive inter-frame connectivity strategy used in 
BM3D-4 is the reason behind 
its significantly better performance. 
Also, using a low-pass 
filter for preprocessing (BM3D-4$_\sigma$) gives BM3D-4
a noteworthy improvement (0.94 dB for the Lena 10-image dataset). 

There can be imprecision in variance stabilisation 
when the VST is applied on a single noisy 
image \cite{azzari2016IterativePoisson}. 
The imprecision is 
even higher because of the initial averaging step in BM3D-1, 
which modifies the noise distribution. 
This is clearly evident from Table \ref{table1} because 
denoising 5 images is surprisingly better than denoising 10 images 
using BM3D-1. We can avoid such a problem using BM3D-4
as it does not average the initial noisy images but retains the original 
noise model.

BM3D-4 is slightly more than $2L$ times 
(two main steps and the extensive patch-matching) slower 
than single-frame BM3D, for a dataset with $L$ frames. 
Our CUDA implementation of BM3D-4 on an NVIDIA GeForce GTX 1070 device 
takes 1.76 seconds for a dataset sized $256 \times 
256 \times 5$. 


\section{Conclusions and Outlook}
\label{sec:concAndOutlook}
In this work we have evaluated the capability of different 
existing extensions of BM3D to remove multi-frame Poisson noise, 
by combining them with a variance stabilising tranformation. 
Our evaluation revealed that the extension (BM3D-4) which retains the 
original noise model and additionally 
makes use of the complete available information through a 
comprehensive connectivity of 2D and temporal information, 
is the best method for multi-frame Poisson noise 
elimination. Our new idea to preprocess the data using low-pass 
filtering gives significant additional quality improvement to the 
best performing extension BM3D-4. 

In the future, we will explore the application of BM3D-4 
for eliminating Poisson and Poisson-Gaussian mixture noise of 
varying amplitudes in the same dataset. We also plan to use BM3D-4 
for denoising datasets acquired with electron microscopes, by combining 
it with state-of-the-art motion registration algorithms. 

\medskip
{\bf Acknowledgements.}
J.W. has received funding from the European Research Council (ERC)
under the European Union's Horizon 2020 research and innovation programme
(grant agreement no. 741215, ERC Advanced Grant INCOVID).


\bibliographystyle{IEEEbib}
\bibliography{myrefs}
\end{document}